\documentclass[aps,prb,twocolumn,superscriptaddress,a4paper,longbibliography]{revtex4-2}
\usepackage[dvipdfmx]{graphicx}
\usepackage{color}
\usepackage{amsmath}
\usepackage{amssymb}
\usepackage{lipsum}
\begin{document}

\title{Ground State Properties of Quantum Skyrmions described by Neural Network Quantum States}

\author{Ashish Joshi}
\affiliation{Department of Physics, Kyoto University, Kyoto 606-8502, Japan}
\email[]{joshi.ashish.42a@st.kyoto-u.ac.jp}
\author{Robert Peters}
\affiliation{Department of Physics, Kyoto University, Kyoto 606-8502, Japan}
\author{Thore Posske}
\affiliation{I. Institute for Theoretical Physics, Universit\"at Hamburg, Notkestraße 9, 22607 Hamburg, Germany}
\affiliation{The Hamburg Centre for Ultrafast Imaging, Luruper Chaussee 149, 22761 Hamburg, Germany}
\begin{abstract}

We investigate the ground state properties of quantum skyrmions in a ferromagnet using variational Monte Carlo with the neural network quantum state as variational ansatz. We study the ground states of a two-dimensional quantum Heisenberg model in the presence of the Dzyaloshinskii-Moriya interaction (DMI) and show that the ground state accommodates a quantum skyrmion for a large range of parameters, especially at large DMI. The spins in these quantum skyrmions are weakly entangled, and the entanglement increases with decreasing DMI. We also find that the central spin is completely disentangled from the rest of the lattice, establishing a non-destructive way of detecting this type of skyrmion by local magnetization measurements. While  neural networks are well suited to detect weakly entangled skyrmions with large DMI, they struggle to describe skyrmions in the small DMI regime due to nearly degenerate ground states and strong entanglement. In this paper, we propose a method to identify this regime and a technique to alleviate the problem. Finally, we analyze the workings of the neural network and explore its limits by pruning. Our work shows that neural network quantum states can be efficiently used to describe the quantum magnetism of large systems exceeding the size manageable in exact diagonalization by far.
\end{abstract}

\maketitle

\section{Introduction}

Classical magnetic skyrmions are magnetic structures with vortex-like configurations characterized by a quantized skyrmion number. After pioneering theoretical work \cite{BogdanovYablonskii1989,bogdanov_physical_2020}, skyrmions have been discovered in a variety of materials, including MnSi, FeCoSi, FeGe, and others \cite{muhlbauer_skyrmion_2009,yu_real-space_2010,kotani_observation_2018,heinze_spontaneous_2011,nagaosa_topological_2013}, with sizes ranging from micrometers to nanometers. These quasiparticles can be created by competition between the exchange interaction and the anti-symmetric Dzyaloshinskii-Moriya interaction (DMI) \cite{nagaosa_topological_2013}, or by frustration \cite{lohani_quantum_2019}, or dynamically by an electric current \cite{stier_skyrmionanti-skyrmion_2017,everschor-sitte_skyrmion_2017} or by boundary effects \cite{siegl_controlled_2022}. Magnetic skyrmions have potential uses in magnetic storage devices due to their topological protection and ease of motion under electric currents \cite{kiselev_chiral_2011,romming_writing_2013,schutte_inertia_2014,jonietz_spin_2010,yu_skyrmion_2012}. The observation of skyrmions with sizes a few times the atomic lattice spacing raises the question about the importance of quantum effects in these systems, meriting a purely quantum mechanical analysis to study them. 

In the past few years, there have been some works addressing the quantum nature of magnetic skyrmions, with the focus on quantum spin systems in the presence of DMI or frustration \cite{takashima_quantum_2016,gauyacq_model_2019,haller_quantum_2022,siegl_controlled_2022,lohani_quantum_2019,sotnikov_probing_2021,maeland_quantum_2022}, or very recently in systems with itinerant magnetism \cite{Kobayashi2022} and with f-electron systems \cite{peters_quantum_2023}.
Most studies have used exact diagonalization techniques to tackle the problem of quantum skyrmions in spin lattices, which inherently puts a limit on the system size they can consider. Recently, quantum skyrmions on a larger spin lattice were considered by Haller et al.~\cite{haller_quantum_2022}, using density matrix renormalization group (DMRG) methods. This work discovered a skyrmion lattice phase that would not be tractable using exact diagonalization. Although having many successful applications, DMRG becomes numerically challenging in two dimensions due to increasing entanglement with the system sizes described by the area law \cite{schollwock_density-matrix_2005,schollwock_density-matrix_2011}. On the other hand, due to the presence of DMI, quantum Monte Carlo methods suffer from the sign problem, which slows down the optimization process.

In recent years, artificial neural network-based variational methods have been introduced to approximate the quantum many-body problem, achieving results comparable to state-of-the-art methods \cite{carleo_solving_2017,nomura_restricted-boltzmann-machine_2017,choo_symmetries_2018,saito_machine_2018,szabo_neural_2020,hibat-allah_recurrent_2020,kochkov_learning_2021,inui_determinant-free_2021,gao_efficient_2017}. In these variational methods, an artificial neural network is used to represent the variational wave function, known as a neural network quantum state (NQS), which then learns the target state using a gradient-based optimization scheme. NQS-based variational methods offer a novel approach to studying a wide range of quantum many-body systems, especially in two and three dimensions where existing methods involve a high level of complexity. NQSs with various structures have been successfully applied to frustrated spin systems in two dimensions \cite{carleo_solving_2017,szabo_neural_2020,hibat-allah_recurrent_2020,chen_neural_2022,kochkov_learning_2021} and have recently been shown to have the ability to capture long-range quantum entanglement with an expressive capacity greater than conventional methods \cite{sharir_neural_2022}  

In this paper, using NQSs we show that quantum skyrmions (QSs) are the ground states for a wide range of parameters in the two-dimensional spin-$1/2$ Heisenberg Hamiltonian with DMI in a ferromagnetic medium. To study quantum entanglement in this system, we calculate Renyi entropy of second order and demonstrate that the entanglement in the QS ground state decreases with increasing DMI.
Previous work indicated that the central spin of a quantum skyrmion can have vanishing concurrence with its surrounding spins \cite{haller_quantum_2022}. Interestingly, we also find that the central spin in the QS ground state is completely disentangled from the rest of the spins within the error bars of our method. This opens up a way of detecting quantum skyrmions experimentally without destroying their quantum nature. 
While we find stable QSs at large DMI, the variational method is insufficient to learn the ground state wave function at small DMI. 
An analysis of small systems reveals that the variational method finds a superposition of the ground state and the first excited state due to a tiny excitation gap. Motivated by this, we present a projection-based method to improve the variational ground state in this region. 
Finally, we analyze the internal structure of our NQS ansatz by inspecting the trained network weights and pruning. While the lowly entangled NQS does not change significantly upon pruning, the performance degrades rapidly with pruning in the highly entangled NQS. Our work shows that an NQS variational ansatz can be used to efficiently approximate spin systems with medium to high DMI at system sizes out of reach for exact methods.

The paper is organized as follows: In Sec.~\ref{model}, we describe the Heisenberg Hamiltonian on a square lattice in the presence of DMI. In Sec.~\ref{method}, we briefly discuss the variational method, along with the neural network structure, with more details in Appendix A. In Sec.~\ref{groundstate}, we study the different ground states of this system. In Sec.~\ref{quantumentanglement}, we study the entanglement in the ground state by calculating the Renyi entropies. In Sec.~\ref{networkinterpretation}, we obtain and present insights into the workings of the trained NQS. Finally, we conclude the paper in Sec.~\ref{summary}.

\section{Model}
\label{model}
We study the ground state of the two-dimensional spin-$1/2$ Heisenberg Hamiltonian on a square lattice in the presence of the Dzyaloshinskii-Moriya interaction (DMI) and a strong external magnetic field at the boundaries that simulates a ferromagnetic background,
\begin{equation}
    \begin{split}
        H=&-J\sum_{\left<ij \right>}(\sigma_i^x\sigma_j^x+\sigma_i^y\sigma_j^y) - A\sum_{\left<ij \right>}\sigma_i^z\sigma_j^z\\ 
        &-D\sum_{\left<ij \right>}(u_{ij}\times\hat{z})\cdot({\sigma}_i\times{\sigma}_j) + B^z\sum_{b}\sigma^z_b.
    \end{split}
    \label{Ham}
\end{equation}
Here, $J$ is the Heisenberg exchange interaction, $A$ is the Heisenberg anisotropy term, $D$ is the DMI, and $B^z$ is the external magnetic field along the $\hat{z}$-axis acting only on the boundary spins indexed with $b$. The Pauli operator at the $i$-th lattice site is ${\sigma}_i=(\sigma^x_i,\sigma^y_i,\sigma^z_i)$ and $u_{ij}$ is the unit vector pointing from site $i$ to site $j$. We consider $\hbar=1$. The sum in the first three terms is over the nearest neighbor lattice sites, while the last term only covers the boundary sites. 

The Hamiltonian in Eq.~(\ref{Ham}) can be considered the quantum analog of a classical spin model in which the competition between the noncolinear DMI and the ferromagnetic Heisenberg exchange interaction gives rise to the formation of magnetic skyrmions. In classical systems, these skyrmions are often stabilized by an external magnetic field over the whole lattice. However, we only apply the magnetic field ($B^z=10J$) to fix the spins at the boundaries. This is the main difference between the Hamiltonian in our work and that in Ref.~\cite{haller_quantum_2022}, where the authors study a similar system but with a bulk external magnetic field. Thus, our model describes a single quantum skyrmion embedded in a ferromagnetic medium. We leave the study of a quantum skyrmion lattice with NQS and the comparison with DMRG results \cite{haller_quantum_2022} for future works. 

\section{Method}
\label{method}
\begin{figure}
        \centering
        \includegraphics[width=\linewidth]{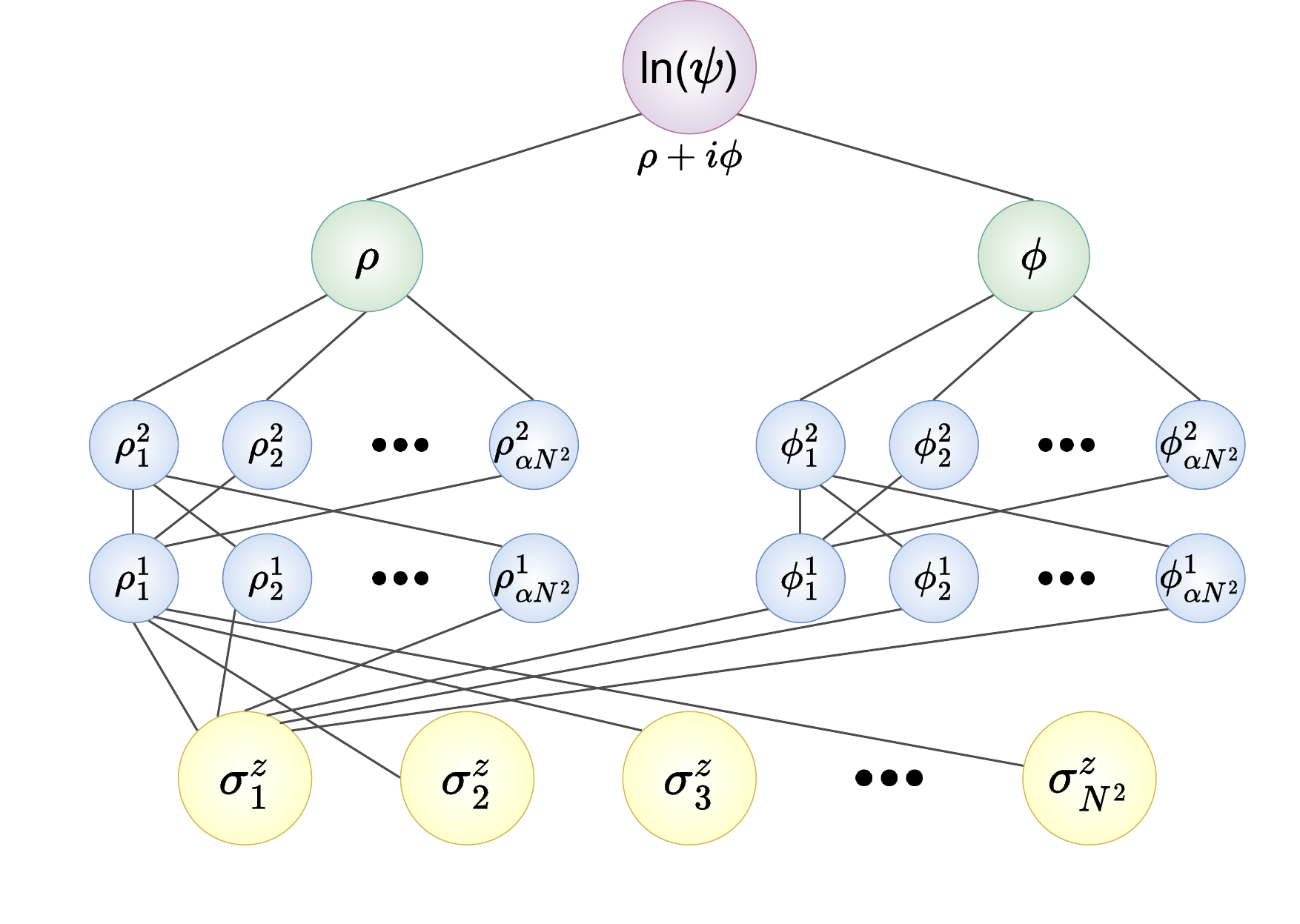}
	\caption{Neural network structure used as NQS. The inputs are the spin configurations in the $\sigma^z$ basis, and the output is the logarithm of the wave function. There are two fully connected networks, with two hidden layers in each, to learn the phase and the amplitude part of the wave function separately. Each hidden layer consists of $\alpha N^2$ neurons.}
	\label{NN}
\end{figure}

The idea behind neural network quantum states is to use the output of an artificial neural network to represent the complex-valued coefficients $\psi_\theta(\sigma)$ in the variational wave function,
\begin{equation}
    \left|\psi_\theta\right>=\sum_{\sigma}\psi_{\theta}(\sigma)\left|\sigma\right>.
\end{equation}
Here, $\left|\sigma\right>$ are  local basis states, which in our case are the eigenstates of the $\sigma^z$ operators, and $\theta$ are the variational parameters of the neural network. In this paper, we use two fully connected feed-forward neural networks to each represent the phase and modulus part of the wave function, see Fig.~\ref{NN}, and take the logarithm of the wave function as the total output \cite{szabo_neural_2020}
\begin{equation}
    \text{ln}\left(\psi_\theta(\sigma)\right)=\mathbf{\rho}(\sigma)+i\mathbf{\phi}(\sigma).
\end{equation}

\begin{figure*}
	\centering
     \begin{minipage}[b]{0.49\linewidth}
        \includegraphics[width=\linewidth]{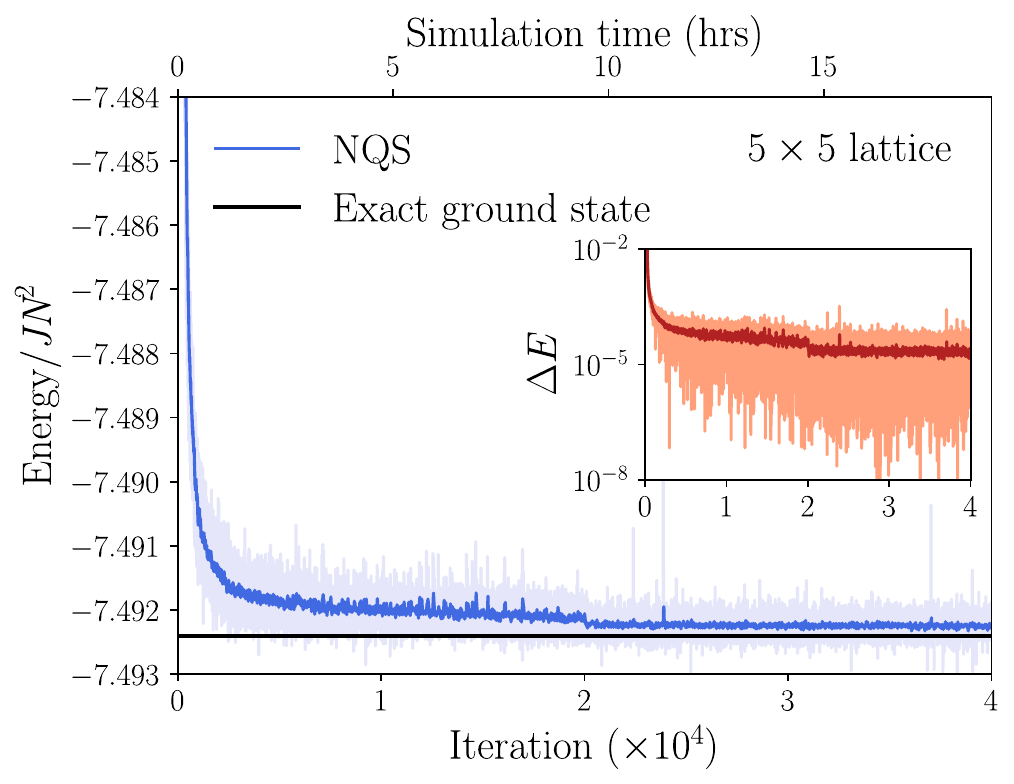}
        (a)
	\end{minipage}	
	\begin{minipage}[b]{0.49\linewidth}
        \includegraphics[width=\linewidth]{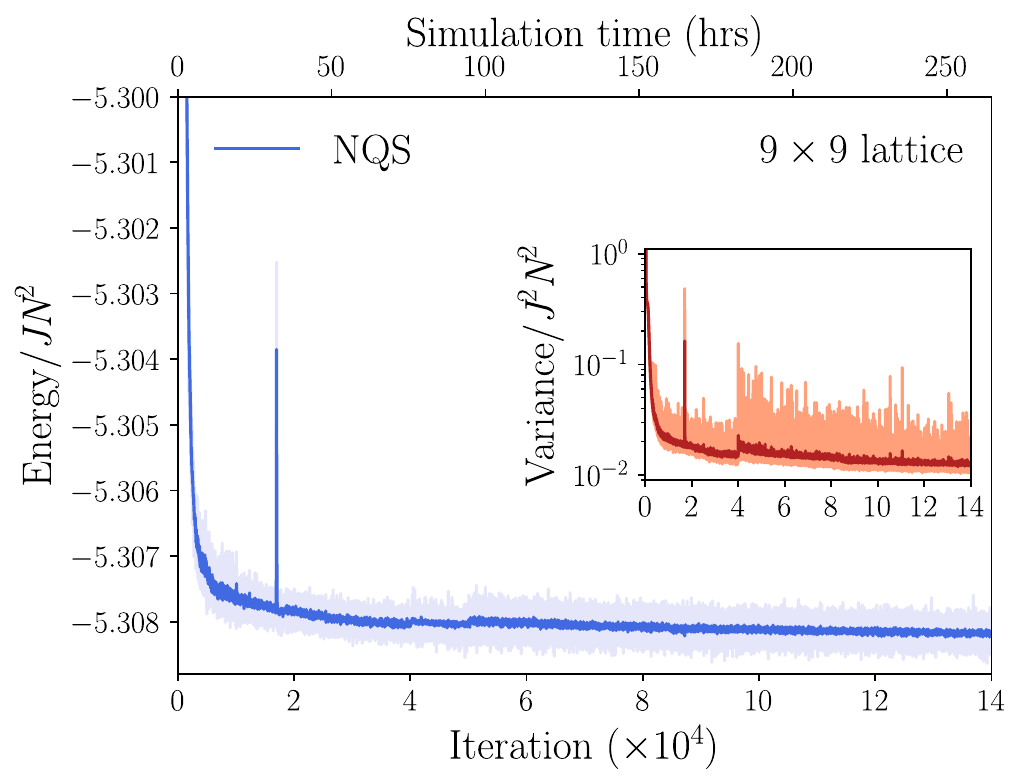}
        (b)
	\end{minipage}
	
	\caption{Convergence of the NQS training procedure: The figure shows the convergence of variational energy per spin to the ground state of a $5\times 5$ lattice (a) and a $9\times 9$ lattice (b) over the number of iterations at the bottom axis and time elapsed at the top axis (see \ref{appendix_A} for hardware specifications). The inset in (a) shows the relative error $\Delta E$ in the ground state energy (see Eq.~(\ref{error})) with respect to the exact ground state energy (black line). The inset in (b) shows the energy variance per spin in dependence on the number of iterations. Light blue and orange lines show the values at each iteration while dark blue and red lines show the moving average over 30 iterations. }
	\label{energy}
\end{figure*}

 The network takes the configuration of the spins on the two-dimensional lattice as the input. Both the phase and the modulus part of the network consist of two fully connected hidden layers with $\alpha N^2$ neurons in each layer, where $N$ is the length of one side of the lattice, and we use $\alpha=2$ in this paper. We use the rectified linear unit (reLU) as the nonlinear activation function. The optimization of the variational wave function is achieved by minimizing the loss function $L_\theta$, i.e., the variational energy, with respect to the variational parameters
 \begin{equation}
    L_\theta = \left<\psi_\theta|H|\psi_\theta\right>.
    \label{loss}
\end{equation}

 The phase part of the network is trained first while keeping the modulus part constant before optimizing the whole network. This method of optimization results in better learning of the sign structure of the ground state wave function, as demonstrated in Ref.~\cite{szabo_neural_2020} and also found by our testing. We use Adam as the optimizer \cite{kingma_adam_2017}. The input samples are generated using the Markov chain Monte Carlo. We use NetKet to implement the NQS and Monte Carlo algorithms \cite{vicentini_netket_2022,carleo_netket_2019,mpi4jax:2021}. Details of the optimization procedure and hyperparameters are given in Appendix \ref{appendix_A}.

\section{Ground State}
\label{groundstate}
\begin{figure*}
	\centering
    \begin{minipage}[b]{\linewidth}
        \includegraphics[width=0.475\linewidth]{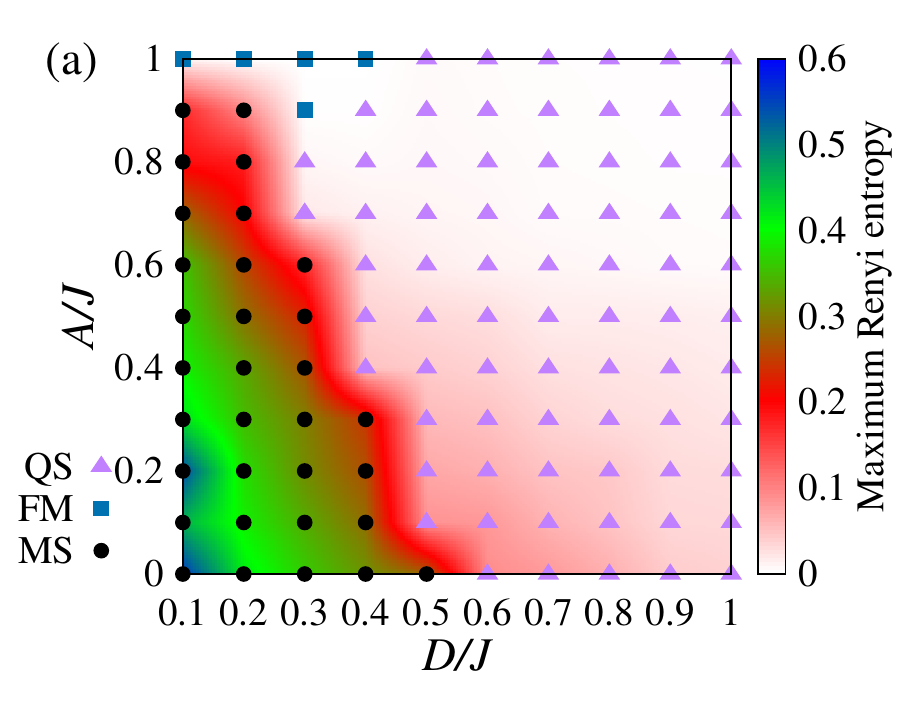}
        \includegraphics[width=0.45\linewidth]{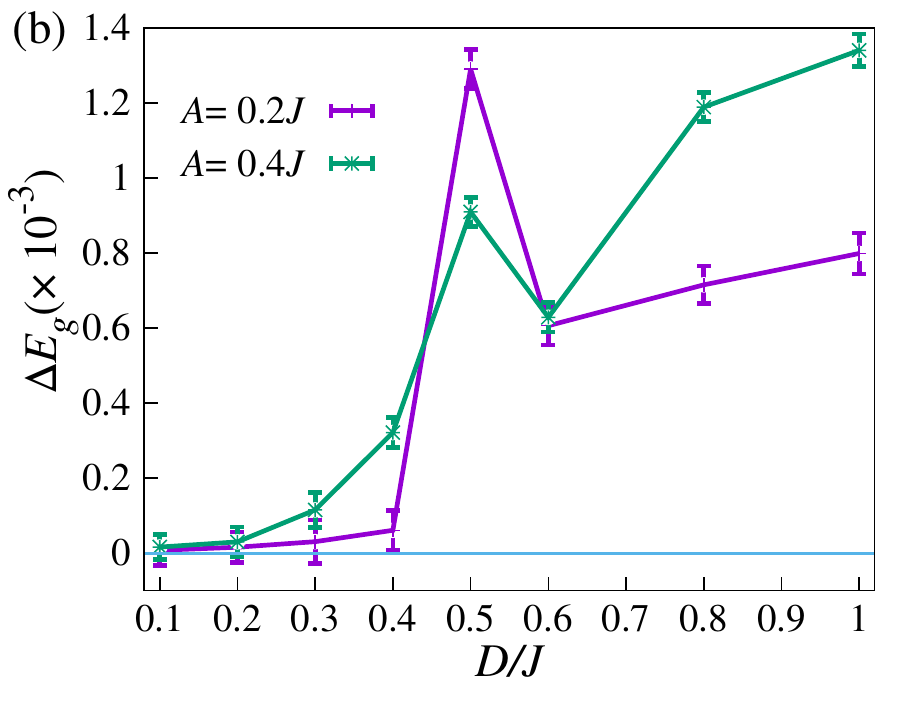}
    \end{minipage}
    \begin{minipage}[b]{\linewidth}
		\includegraphics[width=0.3\linewidth]{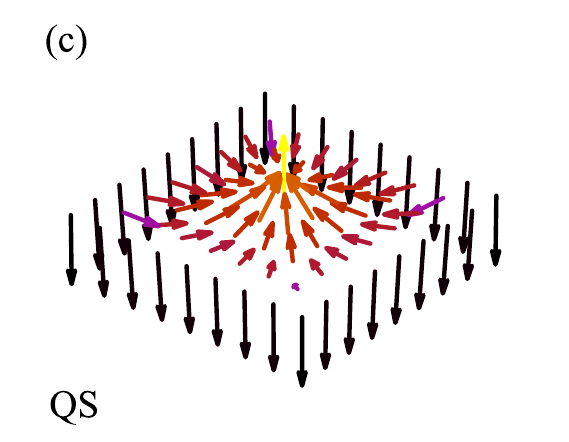}
		\includegraphics[width=0.3\linewidth]{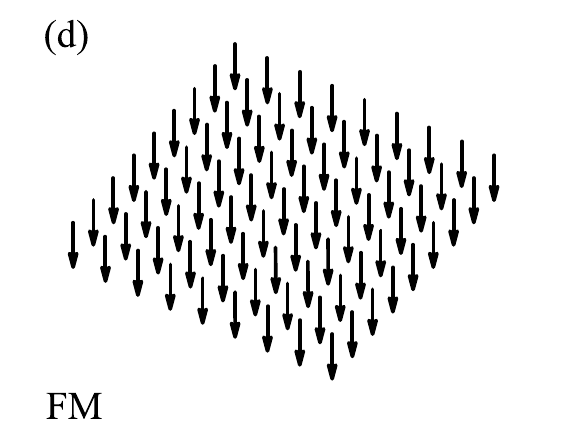}
		\includegraphics[width=0.3\linewidth]{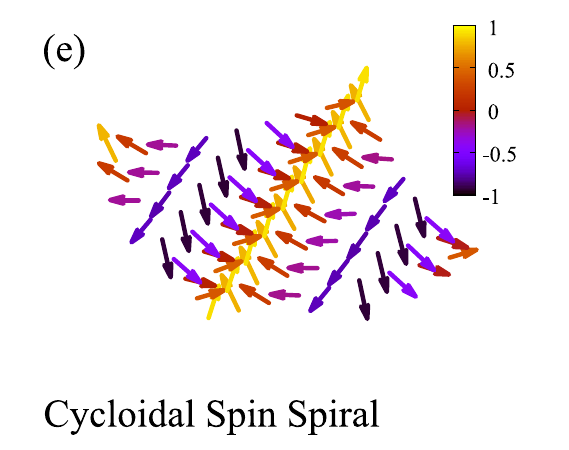}
	\end{minipage}	
	\caption{Ground state diagram and spin expectation values of different ground states of Eq.~(\ref{Ham}) for a $9\times9$ square lattice. (a) Ground state diagram. QS denotes the quantum skyrmion state, FM the ferromagnetic state aligned with the boundary fields, and MS the mixed state (see the main text). The color map shows the maximum Renyi entropy. (b) Relative energy gap $\Delta E_g$ between the ground state and the first excited state found by the neural network quantum state method over DMI at $A=0.2J$ and $A=0.4J$. (c)-(e) Spin expectation values of different ground states. (c) QS at $D=0.5J,A=0.2J$ and (d) FM at $D=0.1J,A=J$. (e) For periodic boundary conditions, we obtain a cycloidal spin spiral instead of a quantum skyrmion at $D=J,A=0.5J$}.
	\label{spinexp}
\end{figure*}

First, we discuss the ability of the NQS ansatz to represent the ground state of the Hamiltonian in Eq.~(\ref{Ham}). To check that our method works correctly, we compare the NQS ground state energy, $E_{NQS}$, for $3\times 3$ and $5\times 5$ spin lattices with the exact ground state energy, $E_{\text{exact}}$, obtained using exact diagonalization. We find that the NQS correctly describes all parameter regimes besides the small DMI regime. While the NQS ground state energies are in agreement with the exact energies within the error margin in this regime, the NQS spin expectation values do not match that of the exact ground state. The reason for this problem lies in an almost degeneracy of the ground state with the first excited state resulting in a significant overlap of the NQS ground state with the excited state found by exact diagonalization. Because of the above, we first present our results for the parameter regime where NQS is accurate and discuss the small DMI regime afterward.

The energy convergence plot for the $5\times 5$ lattice at $D=0.8J$ and $A=0.3J$ is shown in Figure~\ref{energy}(a). Here, the NQS correctly describes the quantum skyrmion ground state. The inset shows the relative error $\Delta E$ in the ground state energy over the number of iterations,
\begin{equation}
    \Delta E=\frac{|E_{\text{NQS}}-E_{\text{exact}}|}{|E_{\text{exact}}|}. 
    \label{error}
\end{equation}
The energy convergence for a $9 \times 9$ lattice over the number of iterations for the Hamiltonian parameters $D=0.5J$ and $A=0.2J$ is shown in Fig.~\ref{energy}(b). 

The ground state diagram for this lattice is shown in Fig.~\ref{spinexp}(a), depending on the DMI, $D$, and the anisotropy, $A$. The quantum skyrmion (QS) is the ground state for a wide range of parameters (triangles in Fig.~\ref{spinexp}(a)), especially at stronger DMI, which favors a noncolinear alignment of the neighboring spins. The spin expectation value at the $i$-th site, $\left<\boldsymbol{S}_i\right>=\left<{\sigma}_i/2\right>$, for the ground state at $D=0.5J$ and $A=0.2J$ is shown in Fig.~\ref{spinexp}(c). A fundamental difference from the case of classical magnetic skyrmions is that the expectation value of the length of the spins, $|\left<\boldsymbol{S}_i\right>|$, is reduced in the QS state. For the ground state in Fig.~\ref{spinexp}(c), $|\left<\boldsymbol{S}_i\right>|$ ranges from $0.92\frac{\hbar}{2}$ in the ring around the center to $1.00\frac{\hbar}{2}$ at the boundary and the center of the quantum skyrmion. Among the QS ground states, the minimum of $|\left<\boldsymbol{S}_i\right>|=0.90\frac{\hbar}{2}$ is found at $D=0.6J$ and $A=0$. The spins are not merely rotated from the boundary to the center, as is the case with classical spins, but are a superposition of the local eigenstates of spin operators in different directions.

For large $A$ and small DMI, the ground state is a ferromagnet (FM) as the spins align in the direction parallel to the boundary fields (squares in Fig.~\ref{spinexp}(a)). An example is shown in Fig.~\ref{spinexp}(d) for $D=0.1J$ and $A=J$. Now, we discuss the parameter regime where the NQS struggles to find the correct ground state. As both DMI and $A$ decrease, the magnitude of the spin expectation values also decreases. We find that in this regime, marked by circles in Fig.~\ref{spinexp}(a), the quantum skyrmion only exists as a metastable state for some parameters, observed in the form of a local minimum during the optimization procedure where the NQS is stuck for some iterations before converging to the ground state. 
The ground state is characterized by almost vanishing spin expectation values aligned along the $x$ or $y$ direction. As mentioned earlier, for small DMI values, the NQS is not able to resolve the nearly degenerate ground state from the first excited state even in smaller lattices.
Hence, we label this regime where our method does not find either a QS or an FM ground state as a ``mixed state" (MS) (circles in Fig.~\ref{spinexp}(a)). 

A conclusion that must be drawn from this result is that energy convergence cannot be taken as the sole measure of accuracy for the variational ground state. To have an additional metric for quantifying the accuracy of our approach, we calculate the gap between the ground state and the first excited state. This is achieved in the variational Monte Carlo scheme by optimizing a second NQS, $\left|\psi_\theta^1\right>$, orthogonal to the ground state NQS, $\left|\psi_\theta^0\right>$, by adding an additional term in the loss function
\begin{equation}
    L_\theta = \left<\psi_\theta^1|H|\psi_\theta^1\right> + J|\left<\psi_\theta^0|\psi_\theta^1\right>|^2.
    \label{gap_loss}
\end{equation}
We calculate the relative energy gap as $\Delta E_g= (E_0-E_1)/E_0$, where $E_0$ and $E_1$ are the energies corresponding to $\left|\psi_\theta^0\right>$ and $\left|\psi_\theta^1\right>$ respectively, and plot it over the DMI in Fig.~\ref{spinexp}(b). For $\Delta E_g < 2\times 10^{-4}$, we do not obtain a QS or FM ground state. This corresponds to the MS region in the parameter space, where quantum skyrmions with very low spin expectation values might exist for some parameters that our method is not able to resolve, as found for small systems by exact diagonalization \cite{siegl_controlled_2022,sotnikov_probing_2021,lohani_quantum_2019}. This suggests that the NQS-based variational methods generally struggle with almost degenerate states. This scenario observed here for quantum spin systems, is well known from finite size electronic topological systems, which only reach exact degeneracy in the thermodynamic limit.

Projection Monte Carlo techniques exist to improve the variational ground state. However, the presence of complex off-diagonal terms in the Hamiltonian makes it difficult to use them stochastically \cite{becca_sorella_2017}. We use an alternative projection method to remove the excited state contributions in the variational ground state (see Appendix~\ref{appendix_B}). After this improvement of the wave function, we obtain the correct ground state for the $3\times3$ lattice but not for the $5\times5$ lattice. Thus, while the NQS is able to represent the correct ground state for all parameters in the case of a $3\times3$ lattice, it is not able to learn it in the small DMI region in our variational Monte Carlo scheme. 

In addition to the spin expectation values, we calculate the skyrmion number $C$ using the normalized spin expectation values, $\textbf{n}_i=\left<\boldsymbol{S}_i\right>/|\left<\boldsymbol{S}_i\right>|$, to define quantum skyrmions \cite{siegl_controlled_2022}
\begin{equation}
    C=\frac{1}{2\pi}\sum_\Delta \text{tan}^{-1}\left(\frac{\textbf{n}_i\cdot(\textbf{n}_j\times\textbf{n}_k)}{1+\textbf{n}_i\cdot\textbf{n}_j+\textbf{n}_j\cdot\textbf{n}_k+\textbf{n}_k\cdot\textbf{n}_i}\right),
    \label{c_no}
\end{equation}
where the sum runs over all elementary triangles $\Delta$ of the triangular tessellation of the quadratic lattice, having the sites $i$, $j$, and $k$ as corners. $C$ gives the number of times the spins wind around a unit sphere and is an integer for quantum skyrmions. In our model, we find $C=1$ for the quantum skyrmion ground state and $C=0$ otherwise. Furthermore, using unnormalized spin expectation values in Eq.~(\ref{c_no}), $\textbf{n}_i=\left<2\boldsymbol{S}_i\right>$, results in a non-integer number $Q$ that indicates the `quantum' nature of skyrmions \cite{siegl_controlled_2022}, similar to other quantum measures \cite{sotnikov_probing_2021}. $Q$ decreases as the entanglement increases and the spin expectation values decrease. For the QS ground states, we find a lower threshold of $Q=0.9$. 

Lastly,  we note that using periodic boundary conditions without ferromagnetic boundaries ($B^z=0$), we do not find a QS ground state. Instead, we obtain a cycloidal spin spiral (Fig.~\ref{spinexp}(e)), which is consistent with DMRG findings \cite{haller_quantum_2022} and the fact that unfrustrated classical skyrmions require a magnetic field for stabilization. 
Here, a QS state minimizes the energy of a finite region of the lattice if the boundary of this region is ferromagnetically ordered. Furthermore, the quantum skyrmion ground state is stable in the presence of an additional bulk magnetic field $B^z_\text{ext}\sum_j\sigma_j^z$ with $B^z_\text{ext}$ up to the order of $2J$ (not shown here), above which the ground state is a ferromagnet aligned along the applied field.

\section{Quantum Entanglement}
\label{quantumentanglement}
Entanglement is an important property of quantum systems that is absent in classical systems. In this section, we investigate whether the spins in the ground state are entangled by calculating the Renyi entropy as a measure of entanglement. The Renyi entropy of the order $\alpha$, where $\alpha\geq0$ and $\alpha\neq1$, is defined as,
\begin{equation}
    S_\alpha(\rho_A)=\frac{1}{1-\alpha}\text{log}(\text{Tr}(\rho^\alpha_A)).
\end{equation}
Here, $\rho_A$ is the reduced density matrix obtained after splitting the system into two regions $A$ and $B$ and tracing out the degrees of freedom in region $B$. The Renyi entropy is a non-negative quantity that is zero for a pure state and takes the maximum value $\text{log}(\text{min}(d_1,d_2))$, where $d_1$ and $d_2$ are the dimensions of the Hilbert space in region $A$ and $B$, respectively. We take region $A$ as a single spin and region $B$ as the rest of the lattice to obtain the entanglement of each spin with its environment. We calculate the $\alpha=2$ Renyi entropy, $S_2(\rho_A)$, using the expectation value of the `Swap' operator \cite{hibat-allah_recurrent_2020,hastings_measuring_2010} (see Appendix~\ref{appendix_C}).

\begin{figure}
	\centering
		\includegraphics[width=\linewidth]{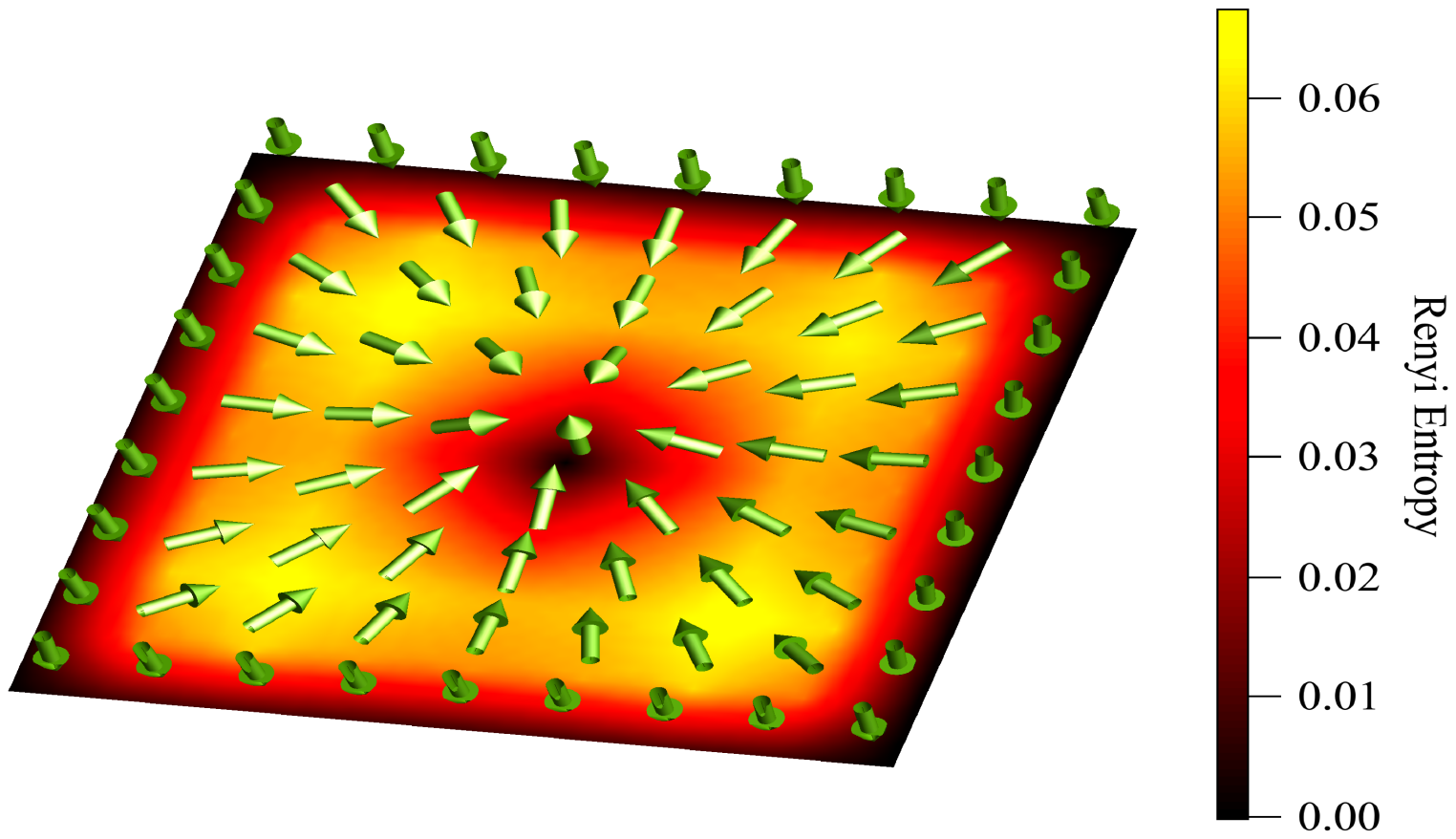}\\
	\caption{Renyi entropy of each spin with its environment for the QS at $D=0.5J$ and $A=0.2J$.}
	\label{EE}
\end{figure}

The maximum Renyi entropy associated with the parameters is shown as a heatmap in the ground state diagram in Fig.~\ref{spinexp}(a). In the QS state for large values of DMI, we find $S_2(\rho_A)\approx 0$ irrespective of which spin $A$ we consider, which means that these quantum skyrmions can be approximated as product states. However, as we reduce the DMI, we find that the entanglement among the spins increases, with the maximum reaching $\text{max}_A S_2(\rho_A)=0.09$ at $D=0.6J$ and $A=0.0$ for the most entangled spin. Here, the quantum skyrmion cannot be described as a product state. We plot the Renyi entropy $S_2(\rho_A)$ as a heat map over the QS ground state in Fig.~\ref{EE} for the parameters $D=0.5J$ and $A=0.2J$. As the boundary spins are fixed with a large magnetic field, they are not entangled with the rest of the spins. The entropy first increases and then decreases from the boundary to the center, reaching its maximum between the two. One unexpected feature of this QS state is that the central spin is also disentangled from the surrounding spins, even though there is no external magnetic field acting on this site. We find that the Renyi entropy of the central spin is numerically zero for all quantum skyrmions that we obtain in our analysis; there are no accepted spin configurations during the Monte Carlo integration where the central spin points in the opposite direction than the ferromagnetic environment.
This means that the QS is a product state of the central spin and a superposition of the rest of the spins. The disentangled central spin can be used to detect quantum skyrmions using the central spin magnetization as an observable in measurements without destroying the quantum nature of the skyrmionic state.
We note that our results of the entropy for the QS ground state match with those in \cite{haller_quantum_2022}, in which the authors considered a bulk magnetic field instead of a ferromagnetic boundary. There, the DMRG calculations indicate a vanishing entanglement of the central spin in a quantum skyrmion with the rest of the system for a certain parameter regime. Thus, a disentangled central spin might be a general feature of quantum skyrmions.

In the FM parameter region, the entropy is $S_2(\rho_A)=0$, and these states can be represented as product states of the spins aligned with the boundary fields. Decreasing  $A$ for small DMI, we approach the MS, and the entropy reaches its maximum. Thus, the difficulties in obtaining a correct solution in this parameter region might also be due to the highly entangled spins that have almost vanishing spin expectation values, along with the small energy gap between the eigenstates.

\section{Network Interpretation}
\label{networkinterpretation}
In this final results section, we shift our focus towards interpreting the working and training of the neural network. Understanding how the network learns the target problem is integral to machine learning research and provides insights that cannot be obtained only through the final prediction. However, the interpretation of neural networks is a nontrivial problem, and a large number of neurons in multiple layers, as in the present network shown in Fig.~\ref{NN}, makes it even more challenging. 

For the case of NQS and many-body physics, inspecting the weights of the neural network may offer clues towards understanding the inner workings of the network \cite{carleo_solving_2017,szabo_neural_2020}. To achieve this and to avoid dealing with an unmanageable amount of variational parameters, we study the QS ground state of the $5\times5$ lattice. We also use a smaller, fully connected feed-forward neural network as our variational ansatz, with two hidden layers and each layer consisting of $25$ neurons for the phase and modulus parts, corresponding to $\alpha=1$ in Fig.~\ref{NN}. We then transfer the results of our analysis to the calculation in the $9\times 9$ lattice.

\begin{figure}
	\centering
	\begin{minipage}[b]{\linewidth}
		\includegraphics[width=0.49\linewidth]{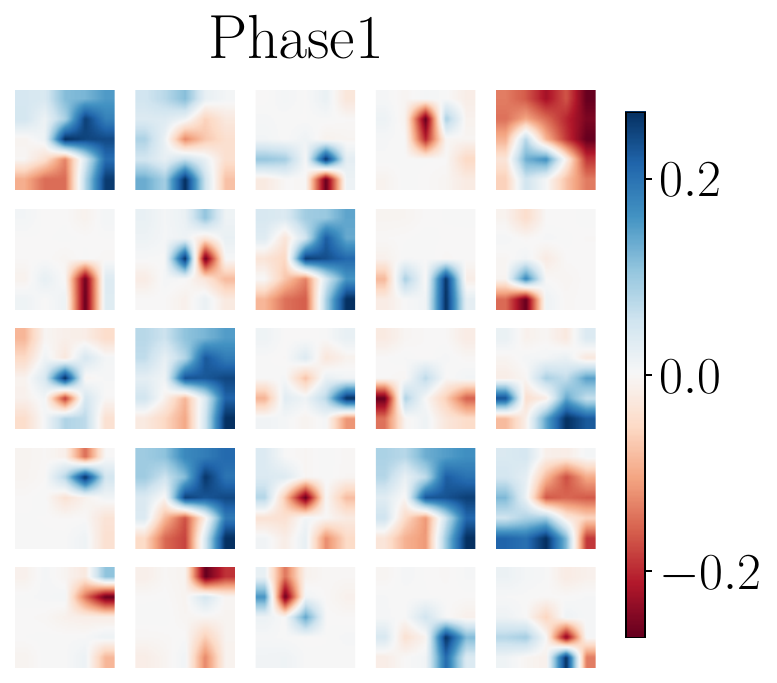}
		\includegraphics[width=0.49\linewidth]{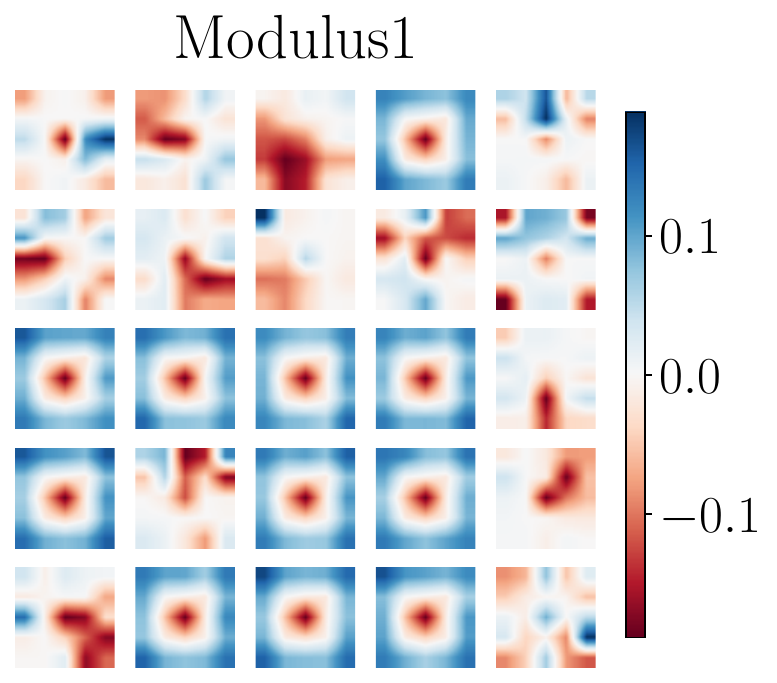}\\
		\includegraphics[width=0.49\linewidth]{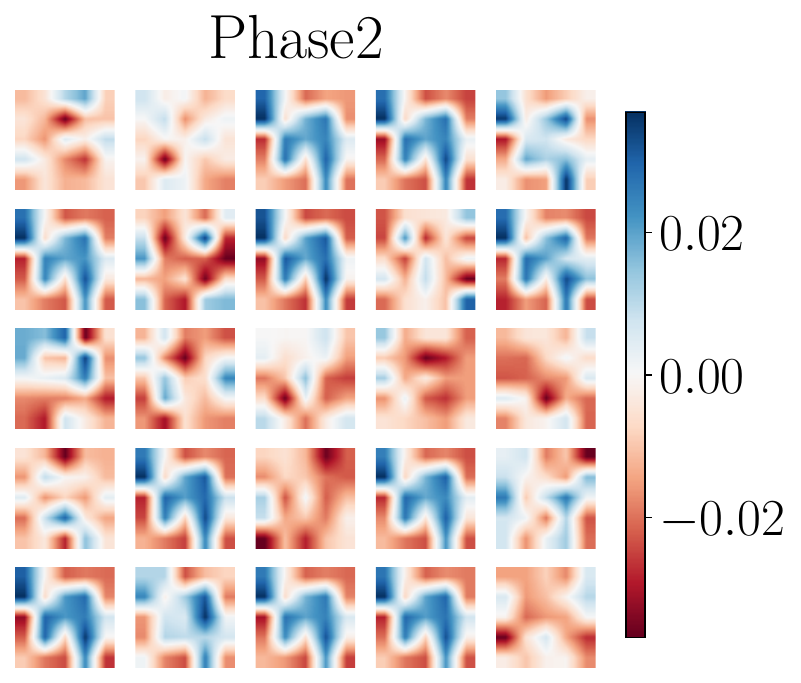}
		\includegraphics[width=0.49\linewidth]{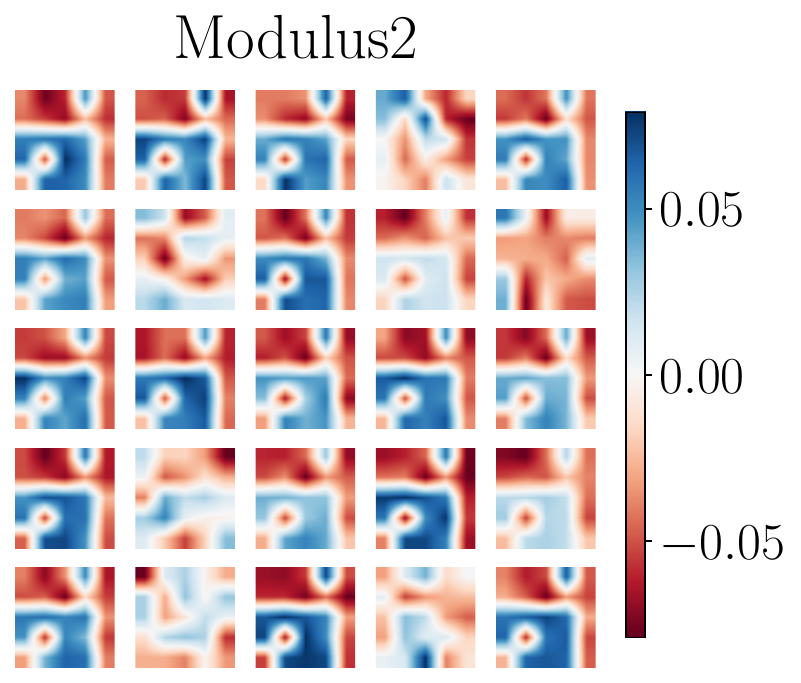}
	\end{minipage}	
	\caption{Weight distribution in the hidden layers of the quantum skyrmion ground state in a $5\times5$ lattice at $D=J$ and $A=0.5J$. Phase1 (Phase2) and Modulus1 (Modulus2) denote the phase and modulus parts in the first (second) hidden layer, respectively. Each block shows the weights inside one hidden neuron. While the first hidden layer learns the essential features of the ground state, most of the neurons in the second hidden layer show a similar pattern.}
	\label{weights_full}
\end{figure}

\begin{figure*}
	\centering
	\begin{minipage}[b]{0.49\linewidth}
		\includegraphics[width=\linewidth]{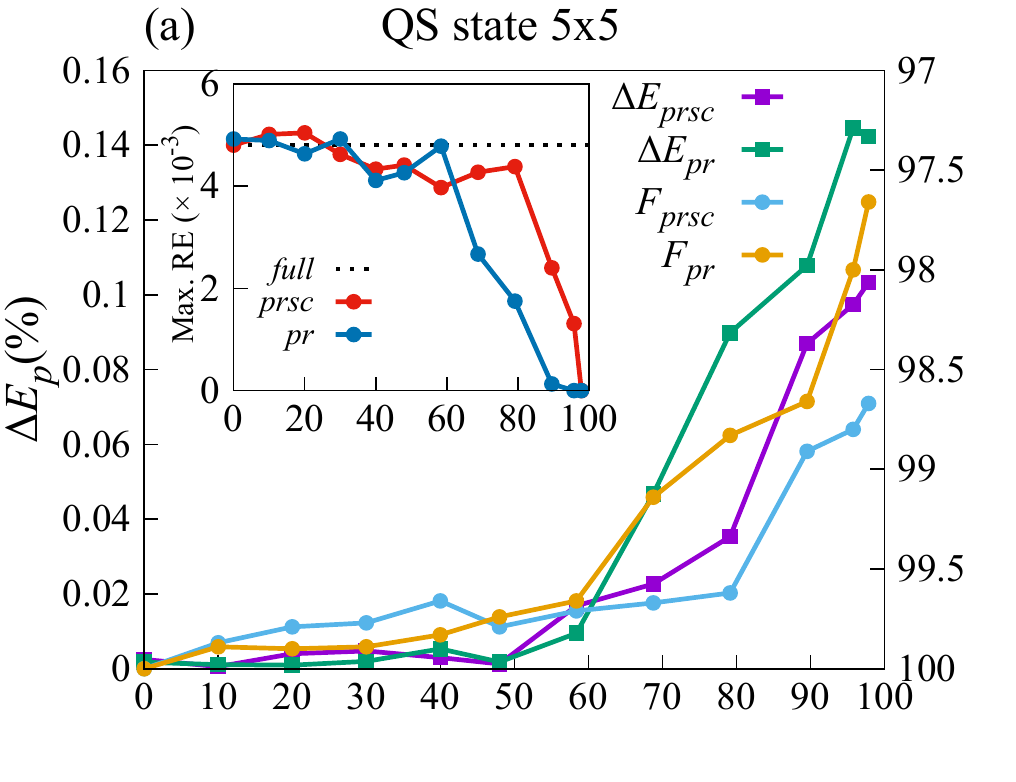}
	\end{minipage}	
	\begin{minipage}[b]{0.49\linewidth}
		\includegraphics[width=\linewidth]{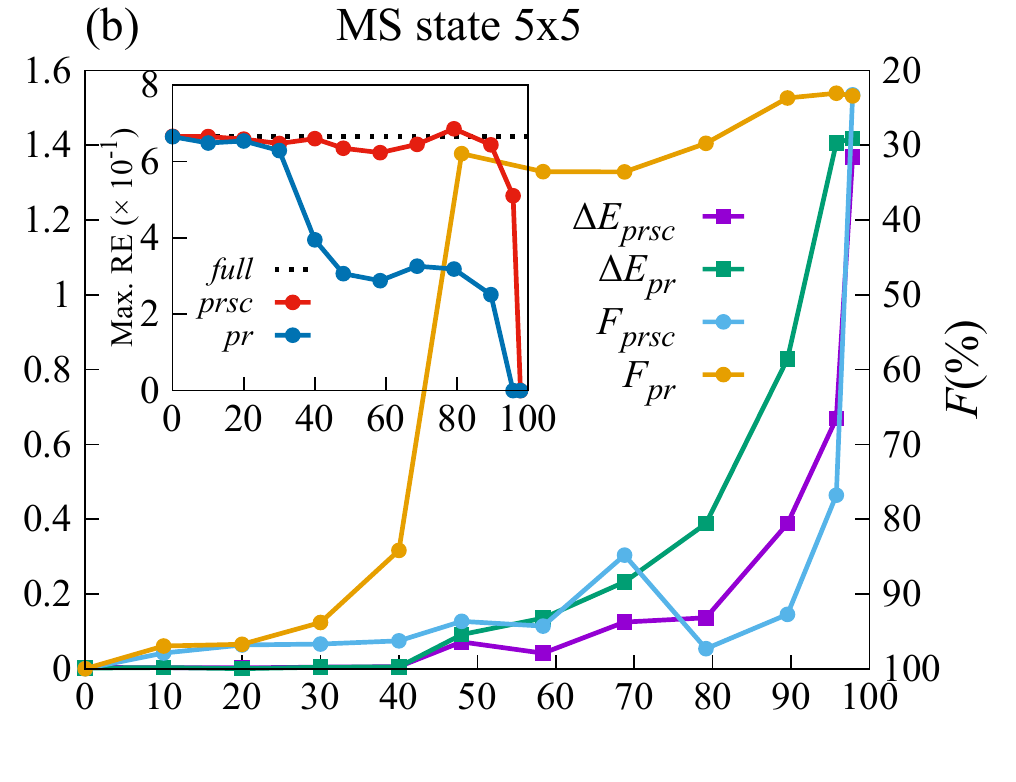}
	\end{minipage}
    \begin{minipage}[b]{0.49\linewidth}
		\includegraphics[width=\linewidth]{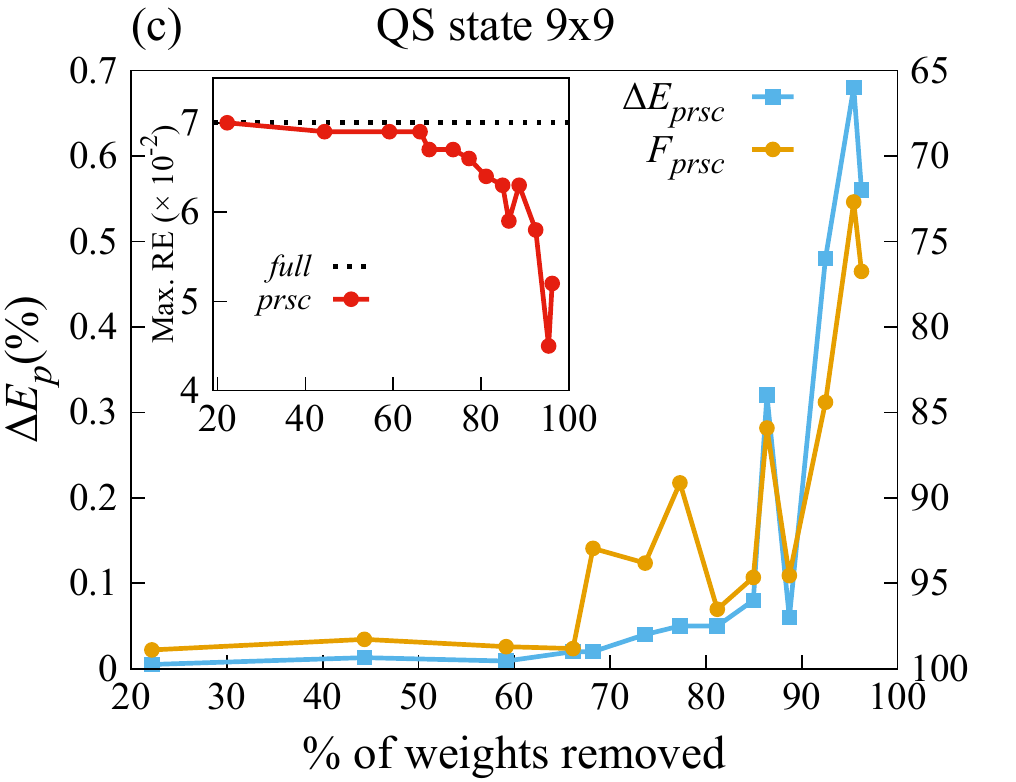}
	\end{minipage}	
	\begin{minipage}[b]{0.49\linewidth}
		\includegraphics[width=\linewidth]{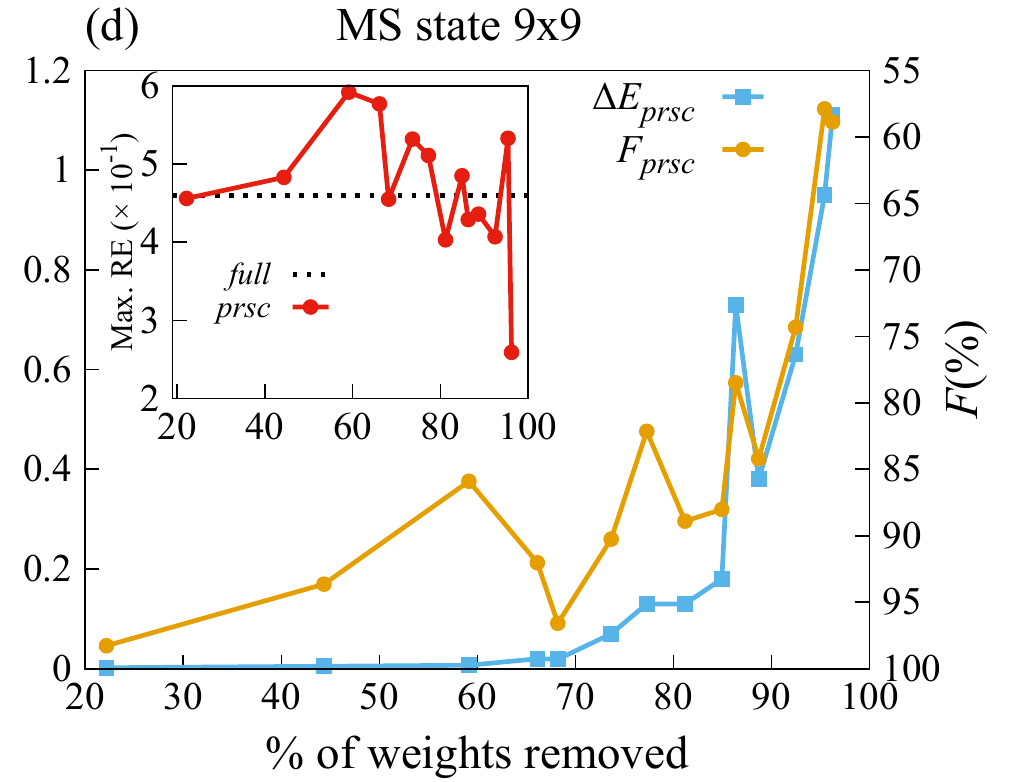}
	\end{minipage}
	\caption{Performance metrics after pruning the neural network. $\Delta E_p$ denotes the relative error in energy and $F$ denotes the fidelity. $pr$ denotes a pruned NQS trained after removing the weights from the full NQS and $prsc$ denotes an identical network to the $pr$ one but trained from scratch. (a) $5\times5$ lattice QS ground state at $D=J$ and $A=0.5J$, (b) $5\times5$ lattice MS ground state at $D=0.1J$ and $A=0.1J$, (c) $9\times9$ lattice QS ground state at $D=0.5J$ and $A=0.2J$, and (d) $9\times9$ lattice MS ground state at $D=0.1J$ and $A=0.1J$.  The inset shows the maximum values of the Renyi entropies.}
	\label{cop}
\end{figure*}
We plot the weights of all neurons of our NQS after training, in a $5\times5$ grid for each layer, in Fig.~\ref{weights_full}. We consider the QS solution at $D=J$ and $A=0.5J$ for our analysis. Inspecting the weights of the first hidden layer, we see that in the phase part, which is trained first to improve the learning of the sign structure of the wave function, each neuron learns a specific part of the wave function. In the modulus part of the first hidden layer, we find that most of the neurons have a skyrmion-like distribution of the weights. This is because the first hidden layer directly takes the spins as inputs; it learns the most important features of the ground state. However, in the second hidden layer, we find that most of the weights in both the phase and the modulus neurons are distributed in a similar pattern and, visually, do not offer a physical interpretation. This raises two questions: first, whether the second hidden layer is essential in the network, and second, whether the neurons with a similar distribution of weights are redundant and can be removed without loss in the accuracy of the network.

In machine learning, pruning is often used to reduce the number of parameters in a neural network to increase computational efficiency without any loss in the accuracy of the network \cite{yang_designing_2017,han_deep_2016,blalock_what_2020}. In most cases, pruning is done post-training by removing the weights with the smallest magnitude and adjusting the remaining weights. After all the pruning steps, only the most important weights are left in the neural network, which can shed some light on the most significant underlying features of the target problem. Pruning could also be important for NQS as a variational ansatz since, with increasing system sizes, the size of the network increases \cite{golubeva_pruning_2022}.

We analyze the effects of pruning to answer the questions we raised above. Again, we consider the $5\times5$ lattice with two types of ground states - the low entanglement QS state at $D=J$,$A=0.5J$ and the high entanglement MS state at $D=0.1J$,$A=0.1J$ in Fig.~\ref{cop}(a)-(b). Starting from the second hidden layer, at each pruning step, 10\% of the neurons from both phase and modulus parts are randomly deleted until only one neuron is left in each of them. Then the same procedure is applied to the first hidden layer. 
After deleting the neurons, the pruned network is trained to adjust the remaining weights ($pr$). Furthermore, a network with the same structure as the pruned network is also trained from scratch ($prsc$) to compare with the $pr$ networks.

As metrics for the performance, we use the relative error, $\Delta E_p$, and the fidelity, $F$, between the original network and the $pr$ or $prsc$ networks \cite{schwandt_quantum_2009},
\begin{align}
    \Delta E_p&=\frac{|E_{\text{full}}-E_{p}|}{|E_{\text{full}}|}, \\
    F&= \left|\left<\psi_{\text{full}}|\psi_{p}\right>\right|^2,
    \label{fidelity_eq}
\end{align}
where $p=pr,prsc$ (see details in Appendix \ref{appendix_A}). In Fig.~\ref{cop}(a)-(b), we plot $\Delta E_p$ and $F$ over the pruning for the $5\times5$ solution, with the maximum Renyi entropies in the insets. 
 For the low entanglement QS solution, the degradation in performance is small even after removing 97\% of the weights, and the fidelity stays over 97\% for both $pr$ and $prsc$ networks. However, for the high entanglement MS solution, the $pr$ and $prsc$ networks show different behavior. The fidelity gradually decreases in the $prsc$ network as the weights are removed. The performance of the pruned network in the high entanglement MS state is worse than in the low entanglement solution. This is expected as it becomes considerably more difficult for fewer neurons to describe the highly entangled state correctly. Moreover, the performance degradation in the $pr$ network is much more severe than in the $prsc$ network. This could be due to the difficulty in leaving the local minimum by the optimizer for the already trained $pr$ network, while $prsc$ networks have the advantage of starting from random weights and thus more flexibility. The maximum Renyi Entropy in both Figs.~\ref{cop}(a) and (b) decreases as the weights are removed. Interestingly, it only becomes zero when only one neuron is left in both hidden layers, showing that NQS can represent entanglement even with a minimal number of neurons. Lastly, we note that on reducing the number of neurons, the optimization process becomes unstable and requires much fine-tuning to converge near the ground state.

In Figs.~\ref{cop}(c) and (d), we show the same results for the $9\times9$ lattice, calculating only $prsc$ networks as they have better performance than the $pr$. The degradation in energy and fidelity, while qualitatively similar to the $5\times5$ case, is more severe. In all four cases, we find that removing neurons from the first hidden layer affects the network's performance more than removing them from the second hidden layer, signifying the importance of the former over the latter. 
This is seen in the very low error in energy until about half of the total weights are removed, after which the error rises drastically. 
Does this mean we can remove the second hidden layer entirely without strongly deteriorating the performance? We find that this is not the case because the performance drastically drops, and the optimization, especially in the high entanglement region, becomes unstable with only one hidden layer. We find that (not shown here) having even a single neuron in the second hidden layer results in greater accuracy than having only one hidden layer with as much as four times the number of neurons. Thus, increasing the width of the network is not the optimal strategy here. On the other hand, having three or more hidden layers makes the optimization process more challenging, and the network is prone to get stuck in a local minimum. Hence, we conclude that the optimal network for our problem should have two hidden layers, with a large number of neurons in the first hidden layer and fewer neurons in the second hidden layer.

\section{Summary}
\label{summary}
In this work, we have studied the ground states of the spin-$1/2$ Heisenberg model in the presence of Dzyaloshinskii-Moriya interaction and Heisenberg anisotropy on a square lattice with ferromagnetic boundaries using variational Monte Carlo. We use a neural network as the variational wave function, with different parts to learn the phase and amplitude of the wave function. We show that a weakly entangled quantum skyrmion ground state, with the skyrmion number $C=1$, exists for a wide range of Hamiltonian parameters. The entanglement increases with decreasing DMI. For large DMI values, a product state can describe the QS ground state. Remarkably, the central spin in the QS state is disentangled from the rest of the spins. 
Furthermore, we analyze the weights of our NQS ansatz and find that while the first hidden layer learns the most important features of the ground state, the second hidden layer is essential to achieve high accuracy. We then test the limits of the NQS by pruning and find that the higher the entanglement, the more deterioration in the performance. 

Finally, we emphasize two of our results: First, our finding that the central spin decouples from the rest of the system and points into the opposite direction than the surrounding ferromagnet can be potentially used as a nondestructive detection scheme for quantum skyrmions by local spin measurements, e.g., by a magnetic scanning tunneling microscope. 
Second, we obtain a region in the parameter space where our method cannot resolve the correct ground state. Instead, we find a superposition between the ground state and the first excited state. This can be traced back to a tiny excitation gap between the ground state and the first excited state and reveals that the NQS ansatz has problems with almost degenerate states, which typically appear in finite size topological systems. While we could devise a scheme to improve the variational state further and separate the ground state from the first excited state in small systems, we could not do this in large spin systems.
Thus, while NQS-based variational methods offer an effective tool to study the quantum skyrmion systems at medium to large DMI, they struggle in the small DMI regime. It is an open question whether other methods like DMRG also struggle in this regime. Improvement of the learning algorithm for NQS-based methods and its comparison with established methods will be our focus for future works.

\begin{acknowledgements}
A. J. is supported by the MEXT Scholarship and Graduate School of Science, Kyoto University under Ginfu Fund. A.J. also acknowledges the funding towards this work from the Kyoto University - University of Hamburg fund. R.P. is supported by JSPS KAKENHI No.~JP23K03300. T.P. acknowledges funding by the Deutsche Forschungsgemeinschaft (project no. 420120155) and the European Union (ERC, QUANTWIST, project no. 101039098). Views and opinions expressed are however those of the authors only and do not necessarily reflect those of the European Union or the European Research Council. Parts of the numerical simulations in this work have been done using the facilities of the Supercomputer Center at the
Institute for Solid State Physics, the University of Tokyo.
\end{acknowledgements}

\appendix
\section{Optimization procedure}
\label{appendix_A}
Given a variational wave function $\left|\psi_\theta\right>$, the expectation value of an operator $O$ can be calculated as \cite{carleo_solving_2017,becca_sorella_2017_vmc},
\begin{align}
        \left<O\right>&=\frac{\left<\psi_\theta\right|O\left|\psi_\theta\right>}{\left<\psi_\theta|\psi_\theta\right>}\nonumber\\        &=\frac{\sum_{\sigma,\sigma'}\left<\psi_\theta|\sigma\right>\left<\sigma|{O}|\sigma'\right>\left<\sigma'|\psi_\theta\right>}{\sum_{\sigma}\left|\psi_\theta(\sigma)\right|^2}\nonumber\\        &=\frac{\sum_{\sigma}\left|\psi_\theta(\sigma)\right|^2\sum_{\sigma'}\left<\sigma|{O}|\sigma'\right>\frac{\psi_\theta(\sigma')}{\psi_\theta(\sigma)}}{\sum_{\sigma}\left|\psi_\theta(\sigma)\right|^2}\nonumber\\
        &=\sum_\sigma p_\theta(\sigma)O_\theta^\text{loc}(\sigma),
        \label{expectation}
\end{align}
where $\psi_\theta(\sigma)=\left<\sigma|\psi_\theta\right>$ and
\begin{align}
    p_\theta(\sigma)&=\frac{\left|\psi_\theta(\sigma)\right|^2}{\sum_{\sigma}\left|\psi_\theta(\sigma)\right|^2},\\
    O_\theta^\text{loc}(\sigma) &= \sum_{\sigma'}\left<\sigma|{O}|\sigma'\right>\frac{\psi_\theta(\sigma')}{\psi_\theta(\sigma)}.
\end{align}
Here, $O_\theta^\text{loc}(\sigma)$ is the local estimator and $p_\theta(\sigma)$ is the probability distribution of $\left|\sigma\right>$. Thus, the quantum expectation value of an observable ${O}$ is the average of a random variable $O_\theta^\text{loc}(\sigma)$ over the probability distribution $p_\theta(\sigma)$. Since the sum over all the states $\left|\sigma\right>$ scales exponentially with the system size, Markov Chain Monte Carlo, with Metropolis-Hastings algorithm, is used to sample a series of states $\left|\sigma\right>_{n}$ and stochastically estimate the expectation values
\begin{equation}
    \left<{O}\right>\approx\frac{1}{N}\sum_{n=1}^{N}O_\theta^{\text{loc}}(\sigma_n),
\end{equation}
where $N$ is the total number of samples. 

The energy of the system can be calculated by taking ${O}$ to be the Hamiltonian. The NQS is then optimized for the ground state iteratively by minimizing the energy using a gradient descent algorithm. Here, we use the Adam optimizer, with the moments $\beta_1=0.9$, and $\beta_2=0.999$ \cite{kingma_adam_2017}. The learning rate $\eta$ is set to $\eta=0.001$ for the phase part and increases linearly from $0$ to $0.001$ over the first $5000$ iterations for the modulus part of the NQS. The learning rate is then reduced to $\eta=0.0001$ after some iterations, evident by a small kink in the energy convergence plots near $20000$ iterations in Fig.~\ref{energy}(a) and near $40000$ iterations in Fig.~\ref{energy}(b). We also tried a stochastic gradient descent optimizer with a stochastic reconfiguration method as a preconditioner to the gradient \cite{carleo_solving_2017} and obtained similar results as with the Adam optimizer but with increased computational cost. A critical step in the optimization procedure is to first optimize the phase part of the network and keep the modulus part constant to facilitate the learning of the phase of the wave function. According to the variational principle, the variational energy is bounded from below by the actual ground state energy, which makes energy a convenient loss function to minimize. We sample by flipping a spin locally $N$ times, each at a random location, where $N$ is the total number of spins in the lattice. This makes one Monte Carlo sweep. We use $10^4$ samples for energy calculation and $10^7$ for all the other expectation values. The calculations were performed on a Xeon Gold 6338 CPU with multi-threading up to 4 cores.

To calculate the fidelity between two NQSs, $\left|\psi_1\right>$ and $\left|\psi_2\right>$ (dropping the dependence on $\theta$ for clarity), we follow a similar procedure as in Eq.~(\ref{expectation}),
\begin{align}
        F&= \frac{\left|\left<\psi_1|\psi_2\right>\right|^2}{\left<\psi_1|\psi_1\right>\left<\psi_2|\psi_2\right>}\nonumber\\        &=\frac{\sum_{\sigma,\sigma'}\left<\psi_1|\sigma\right>\left<\sigma|\psi_2\right>\left<\psi_2|\sigma'\right>\left<\sigma'|\psi_1\right>}{\sum_{\sigma}\left|\psi_1(\sigma)\right|^2\sum_{\sigma'}\left|\psi_2(\sigma')\right|^2}\nonumber\\
        &=\sum_\sigma\frac{\left|\psi_1(\sigma)\right|^2}{\sum_\sigma\left|\psi_1(\sigma)\right|^2}\frac{\psi_2(\sigma)}{\psi_1(\sigma)}\sum_{\sigma'}\frac{\left|\psi_2(\sigma')\right|^2}{\sum_{\sigma'}\left|\psi_2(\sigma')\right|^2}\frac{\psi_1(\sigma')}{\psi_2(\sigma')}\nonumber\\
        &=\sum_\sigma p_1(\sigma)\frac{\psi_2(\sigma)}{\psi_1(\sigma)}\sum_{\sigma'} p_2(\sigma')\frac{\psi_1(\sigma')}{\psi_2(\sigma')}\nonumber.
\end{align}
Thus, $F$ can be evaluated by first sampling from two different probability distributions corresponding to the two NQSs, and then computing the ratio of the wave function amplitudes. 

\section{An alternative projection method}
\label{appendix_B}
Variational wave functions can be improved by projection techniques, which require the variational state to have a finite overlap with the exact ground state. Then, the high-energy components can be projected out by applying the ``power method". However, this method can be done exactly only for systems manageable by exact diagonalization. In other cases, stochastic methods have to be used, requiring the Hamiltonian's off-diagonal terms to be real and non-negative. When this condition is not fulfilled, as in our case with Eq.~(\ref{Ham}) with complex off-diagonal terms, there is a fixed-node approximation for Hamiltonians with real and negative off-diagonal terms and its modification fixed-phase approximation for complex off-diagonal terms \cite{becca_sorella_2017}. 

Here, we propose another method to filter out high-energy components by projecting the Hamiltonian on a few low-energy states, which can be directly obtained by the variational Monte Carlo scheme introduced in the main text.

Given a Hamiltonian ${H}$, its eigenvalue equation is
\begin{equation}
    {H}\left|\phi\right>=E\left|\phi\right>,
\end{equation}
where $E$ and $\left|\phi\right>$ are the eigenvalues and eigenvectors, respectively. By expanding this equation in a complete but not necessarily orthonormal basis $\left|n\right>$, we obtain the generalized eigenvalue equation
\begin{equation}
    \frac{1}{\Omega}\left(\sum_{n_i,n_j} \left<n_j|{H}|n_i\right>\left<n_i|\phi\right> - E\left<n_j|n_i\right>\left<n_i|\phi\right>\right)=0.
    \label{gep}
\end{equation}
Using an incomplete set of states, $\left|n_i\right>$, we can define the projection of the Hamiltonian into the space spanned by these states as  ${H_\textbf{proj}}=\left<n_j|{H}|n_i\right>$ and the overlap matrix ${X}=\left<n_j|n_i\right>$. If $|n_i\rangle$ are approximations of the ground state and the lowest excited states of the Hamiltonian, the ground state of the projected Hamiltonian will be an improved version of the variational ground state of the full Hamiltonian.

In the converged variational NQS ground state $\left|n_0\right>$, the main component is the exact ground state with small contributions from the excited states. By optimizing a second NQS, which is nearly orthogonal to the ground state NQS, using the cost function
\begin{equation}
    L_\theta = \left<n_1|H|n_1\right> + J|\left<n_0|n_1\right>|^2,\label{cost_function_orthogonal}
\end{equation}
as described in the main text in Eq.~(\ref{gap_loss}), the first excited state can be approximated as $\left|n_1\right>$. This procedure can be repeated to approximate the excited states of ${H}$. 
We then can use these variational low-energy states to calculate the projected Hamiltonian and the overlap matrix in a Markov Chain Monte Carlo scheme. We note that even by using the cost function Eq.~(\ref{cost_function_orthogonal}), there is no guaranty that the overlap of $|n_0\rangle$ and $|n_1\rangle$ exactly vanishes.
We use a similar procedure as in Eq.~(\ref{expectation}) to calculate the matrix elements of the projected Hamiltonian and the overlap matrix. 
For the projection on two low-energy states, we sample using the product of these two wave functions $|n_0(\sigma)||n_1(\sigma)|$, as it gave us the best results. The projected Hamiltonian and the overlap matrix are then given as
\begin{eqnarray}
\frac{\left<n_i|H|n_j\right>}{\Omega}&=&\frac{\sum_{\sigma\sigma^\prime}\frac{\vert n_0(\sigma)\vert\vert n_1(\sigma)\vert}{\vert n_0(\sigma)\vert\vert n_1(\sigma)\vert}n^\star_i(\sigma^\prime)n_j(\sigma)\left<\sigma|H|\sigma^\prime\right>}{\sum_\sigma \vert n_0(\sigma)\vert\vert n_1(\sigma)\vert }\nonumber\\
\\
\frac{\left<n_i|n_j\right>}{\Omega}&=&\frac{\sum_{\sigma}\frac{\vert n_0(\sigma)\vert\vert n_1(\sigma)\vert}{\vert n_0(\sigma)\vert\vert n_1(\sigma)\vert }n^\star_i(\sigma)n_j(\sigma)}{\sum_\sigma \vert n_0(\sigma)\vert\vert n_1(\sigma)\vert }
\end{eqnarray}
which determines the normalization constant in Eq.~(\ref{gep}) as $\Omega=\sum_\sigma \vert n_0(\sigma)\vert\vert n_1(\sigma)\vert $.
The wave functions in Eq.~(\ref{gep}) do not need to be normalized because the overlap matrix ${X}$ takes care of any factors arising due to the absence of normalization. Hence, this method can be used in the variational Monte Carlo scheme, which usually considers unnormalized wave functions.

Then, by solving the generalized eigenvalue problem, Eq.~(\ref{gep}), we can filter out the high energy components from the variational ground state. The new variational ground state is $\left|n_0\right>_\text{new}=\sum_{i}\phi_{0i}\left|n_i\right>$, where $\phi_{0i}$ are the components of the lowest energy eigenvector of Eq.~(\ref{gep}).
This procedure is feasible when only a few excited states are mixed in the approximation of the ground state, as the calculation of the excited state itself is variational, and the errors build up with each excited state calculation. This method works well for the $3\times3$ lattice over the entire parameter range, as the variational ground state has negligible overlap with the second and higher excited states. Then, only the calculation of the first variational excited state is required.
However, while it improves the variational energy slightly for larger lattices,  we do not obtain the correct ground state in the small DMI and $A$ region of the ground state diagram. 

\section{Renyi Entropy}
\label{appendix_C}
When a system is divided into two parts, $A$ and $B$, the variational wave function can be written as
\begin{equation}
   \left|\psi_\theta\right>=\sum_{\sigma_A\sigma_B}\psi_\theta(\sigma_A\sigma_B)\left|\sigma_A\right>\left|\sigma_B\right>,
\end{equation}
where $\sigma_A$ and $\sigma_B$ are the basis states in region $A$ and region $B$, respectively. The Renyi entropy of order $\alpha$ between $A$ and $B$ is
\begin{equation}
    S_\alpha(\rho_A)=\frac{1}{1-\alpha}\text{log}(\text{Tr}(\rho^\alpha_A)),
    \label{renyientropy}
\end{equation}
where $\rho^\alpha_A$ is the reduced density matrix obtained after tracing out the degrees of freedom in region B. To calculate the Renyi entropy of the second order ($\alpha=2$), we use the replica trick to evaluate the expectation value of the `Swap' operator on two copies of the variational wave function. The Swap operator swaps the spins in one region with that of another region between the two wave functions \cite{hastings_measuring_2010}
\begin{align}                                                               &\text{Swap}_A\left|\psi_\theta\right>\otimes\left|\psi_\theta\right>=\text{Swap}_A\left(\sum_{\sigma_A\sigma_B}\psi_\theta(\sigma_A\sigma_B)\left|\sigma_A\right>\left|\sigma_B\right>\right)\nonumber\\
&\otimes\left(\sum_{\sigma_A'\sigma_B'}\psi_\theta(\sigma_A'\sigma_B')\left|\sigma_A'\right>\left|\sigma_B'\right>\right)\nonumber\\
&=\sum_{\sigma_A\sigma_B}\psi_\theta(\sigma_A\sigma_B)\sum_{\sigma_A'\sigma_B'}\psi_\theta(\sigma_A'\sigma_B')\left|\sigma_A'\right>\left|\sigma_B\right>\otimes\left|\sigma_A\right>\left|\sigma_B'\right>,
\end{align}
where $\sigma$ and $\sigma'$ are the basis states for the two copies of the wave function. The expectation value of $\text{Swap}_A$ is then given by,
\begin{align}    &\left<\text{Swap}_A\right>=\frac{\left<\psi_\theta\otimes\psi_\theta\right|\text{Swap}_A\left|\psi_\theta\otimes\psi_\theta\right>}{\left<\psi_\theta\otimes\psi_\theta|\psi_\theta\otimes\psi_\theta\right>}\nonumber\\
&=\frac{\sum_{\sigma_A\sigma_B\sigma_A'\sigma_B'}\psi^*_\theta(\sigma_A\sigma_B)\psi^*_\theta(\sigma_A'\sigma_B')\psi_\theta(\sigma_A'\sigma_B)\psi_\theta(\sigma_A\sigma_B')}{\sum_{\sigma\sigma'}\left|\left<\psi_\theta\otimes\psi_\theta|\sigma\otimes\sigma'\right>\right|^2}\nonumber\\
&=\text{Tr}(\rho_A^2)\nonumber\\
&=\text{exp}(-S_2(\rho_A)).
\label{swap_exp}
\end{align}

For the final step we use the definition in Eq.~(\ref{renyientropy}) with $\alpha=2$. In the Monte Carlo scheme, Eq.~(\ref{swap_exp}) can be evaluated as
\begin{align}
    &\left<\text{Swap}_A\right>\nonumber\\ &=\sum_{\sigma_A\sigma_B\sigma_A'\sigma_B'}\frac{\left|\psi_\theta(\sigma_A\sigma_B)\right|^2}{\sum_{\sigma}\left|\psi_\theta(\sigma)\right|^2}\frac{\left|\psi_\theta(\sigma_A'\sigma_B')\right|^2}{\sum_{\sigma'}\left|\psi_\theta(\sigma')\right|^2}\cdot\nonumber\\
    &\quad\quad\quad\quad\quad\quad\quad\frac{\psi_\theta(\sigma_A'\sigma_B)\psi_\theta(\sigma_A\sigma_B')}{\psi_\theta(\sigma_A\sigma_B)\psi_\theta(\sigma_A'\sigma_B')}\nonumber\\    &=\sum_{\sigma_A\sigma_B\sigma_A'\sigma_B'}p_\theta(\sigma)p_\theta(\sigma')\frac{\psi_\theta(\sigma_A'\sigma_B)\psi_\theta(\sigma_A\sigma_B')}{\psi_\theta(\sigma_A\sigma_B)\psi_\theta(\sigma_A'\sigma_B')}.\nonumber\\
    \label{swap_mc}
\end{align}
\bibliography{references}

\end{document}